\documentclass[12pt]{article}
\usepackage{a4,amsmath,amsxtra,amssymb}

\begin{document}

\baselineskip=1.5\baselineskip

\centerline{{\bf Introductory comments to the Archive version of
the paper}}
\centerline{{\bf ``Can We Detect Tachyons Now?''}}

The present paper is placed in this Archive to call attention
of a large community of experimenters to the fact that there exists
a sound theoretical prediction that the light barrier may be
overcome in relatively simple experiments. The paper acquaints the
reader with the situation in an abbreviated and general form, and
outlines the main empiric conditions
that should exist or be created in the considered
case. It should be emphasized that the mentioned prediction
closely relates to the theoretical prediction that gravitational
waves exist, since in terms of general relativity these two
predictions belong to the same family.
Let us also note that massive and very expensive
experiments to search for these waves have already been performed,
some others are being prepared, and new ones are planned.

\vskip64pt

\centerline{{\bf Informative remarks}}

The present version slightly differs from the published one
[Acta Physica Polonica {\bf B31} (2000) 523]. Namely, we have here
small modifications (on p.~9) and extensions (on pp. 2, 3, and 9),
which are marked by underlining. Note that the title page of the
paper carries here number 2.

\newpage

\centerline{{\large CAN WE DETECT TACHYONS NOW?}\footnote[0]{Presented
at the XXVI Mazurian Lakes School of Physics, Krzy\.ze, Poland,
September 1--11, 1999. This contribution is a fragment of the
presentation. The full text can be found in the LANL Archives
(http://xxx.lanl.gov/hep-ph/9911441), \underline{ or in Acta Physica
Slovaca {\bf 50} (2000) 381--395}.}}

\vskip36pt

\centerline{J. K. KOWALCZY\'NSKI}

\vskip12pt

\centerline{Institute of Physics, Polish Academy of Sciences,}

\centerline{Al. Lotnik\'ow 32/46, 02--668 Warsaw, Poland}

\centerline{e-mail: jkowal@ifpan.edu.pl}

\vskip84pt

An exact solution of the Einstein--Maxwell equations enables us to
construct a hypothesis on the production of tachyons.
The hypothesis determines the kinematical relations for the
produced tachyon. It also makes possible to estimate the empiric
conditions necessary for the production. These conditions
can occur when nonpositive subatomic particles of high energy strike
atomic nuclei other than the proton.
This suggests how experiments to search for tachyons can be
performed. According to the hypothesis properly designed experiments
with air showers or with the use of the strongest colliders may be
successful. Failure of the air shower experiments performed
hitherto is explained on the grounds of the hypothesis.

\vskip12pt


\newpage

\centerline{{{\bf\large 1. Introduction}}}


The subject of this note is a hypothesis on the tachyon creation,
based on an exact solution of the Einstein--Maxwell equations.
Details are given in Ref.~[1], where this solution is referred
to as~${\Omega}_{1}$. The solution yields a
realistic picture of the tachyonic phenomenon. This fact
can therefore be regarded as an indication on the part of
general relativity that the tachyon exists in nature,
considering the analogy to many theoretical predictions that found
later empirical confirmation. Our solution invariantly determines
a spacetime point (event) that can
be interpreted as a point of creation of the tachyon; and this
makes the construction of the hypothesis possible. According to our
solution the tachyon produces an electromagnetic field bounded
by the tachyon's shock wave. In the generated spacetime the
gravitational field, i.e. the direct cause of spacetime curvature,
does not exist autonomously but is only an ``emanation'' of
the electromagnetic field. Thus,
even if the latter field were by many orders of magnitude
stronger than the strongest electromagnetic fields observed so far,
the spacetime curvature would be completely negligible. This and the
fact that the tachyon's shock wave is electromagnetic
mean that our solution is proper to
describe an {\it ionizing tachyon belonging to the microworld}.
The hypothesis is presented in Section~2.

Various experimental searches for ionizing tachyons have been
described in a number of papers. A large majority of them is
cited in Refs.
[2--7]. The experiments were of low and high energy type. Failure
of the low energy experiments is explicable by our hypothesis, as
will be seen in Sections 2 and~3. In the high energy experiments
air showers were exploited; and many of the experiments
have reported detection
of tachyon candidates but as statistically insignificant data. A
single possibly positive result [8] has also been rejected~[2].
This situation has presumably disheartened most experimenters (the
last relevant record in the Review of Particle Properties [6] is
dated 1982~[5]), though some efforts were still made~[7].
According to our hypothesis, however, air shower
\underline{(and accelerator)} experiments may
be successful and they are discussed in Section~3.


\centerline{{{\bf\large 2. The hypothesis}}}


The hypothesis says that the tachyon is produced  when  a
neutral subatomic particle of sufficiently  large  rest  mass
(further  called  the generative particle)  is  placed  in
the strong  electromagnetic  field  described just below
(further called the initiating field).
The generative particle is then annihilated giving birth to
the tachyon.

In  this  section  all  the  quantities,   relations,   and
situations are presented in terms of the {\it proper} reference frame
of  the  generative  particle,   with   the use of the  Lorentzian
coordinates $x, y, z$,  and~$t$  (further $t$  does  not   appear
explicitly). We assume that the origin $x = y = z = 0$  of
the  spacelike
coordinates is at the centre of the generative  particle.

Let ${\bf E}$ and ${\bf H}$ be accordingly the electric and magnetic
three-vectors of the initiating field, and let their  components
be denoted by $E_{i}$ and $H_{i}$. There are two types of  the
initiating field:
$$
E_{x} = \mp AB| w| , \hspace {.7cm} E_{z} = \pm 2\hbox{{\it
jABC}}, \hspace {.7cm} H_{y} = \mp \hbox{{\it jAB}}| w| ,
$$
$$
E_{y} = H_{x} = H_{z} = 0,
\eqno (1)$$
\noindent and
$$
E_{y} = \pm AB| w| , \hspace {.7cm}  H_{x} = \mp \hbox{{\it jAB}}|
w| , \hspace {.7cm} H_{z} = \pm 2\hbox{{\it ABC}},
$$
$$
E_{x} = E_{z} = H_{y} = 0,
\eqno (2)$$
\noindent where
$$
j = \pm 1,\qquad | w| \; > 1,\qquad jw < 0,
$$
$$
A > 0, \hspace {.7cm} B := (5w^{2} - 4)^{-1/2} \ge 0, \hspace
{.7cm}  C := (w^{2} - 1)^{1/2} > 0,
\eqno (3)$$
\noindent and where $ w$ is a dimensionless parameter determining
the tachyon's
velocity.  Then,  according  to  the  hypothesis,  the  tachyon
produced in the generative particle  and  fields  (1)--(3)  moves
along a semi-axis~$ z$ with a velocity~$ cw$, $c$ being
the speed of light
in vacuum. The discrete parameter $j$ plays an important  role
in the exact theory based on our solution~[1].
Note that in accordance with the known
properties of the spacelike world lines we may have $| w| = \infty $.

From relations (1)--(3) we see that
$$
{\bf E \perp H},\qquad | {\bf E| \; \neq \; | H|
},\qquad | {\bf E| | H| \;  \neq \; } 0,
\eqno (4)$$
\noindent and that $A = |{\bf E}| > |{\bf H}| $ in the case
(1) and $A = |{\bf H}| > |{\bf E}| $ in  the
case~(2).

Let the tachyon produced in the initiating  field  (1)  and
(3) be called the {\it e-tachyon} (predominance of the electric field)
and that produced in the initiating field (2) and (3) be called the
{\it m-tachyon} (predominance of the magnetic field).  The e-  and
m-tachyons differ since they generate different  electromagnetic
fields.\footnote[1]{On the analogy of the subluminal microworld, in
which only one type of charges (electric) is known, we may suspect
that only one type of our tachyons exists in
nature (i.e. either the e-tachyons or the m-tachyons),
but today we do not yet know which one. Thus, for safety, both
types should be considered.}

Let $U$ be defined as follows: $U = \; | {\bf H| }^{-1}| {\bf E| }$
in the case~(1)
and $U = \; | {\bf E| }^{-1}| {\bf H| }$ in the case~(2).
Thus, by relations (1)--(3), we have $U > 1$ and
$$
U^{2} = 5 - 4w^{-2},
\eqno (5)$$
\noindent i.e.
$$
1 < U^{2} \le  5.
\eqno (6)$$
If the angle between  the  tachyon  path  (a
semi-axis~$ z$) and the longer three-vector of  the  initiating
field is denoted by $\alpha $, then
$$
\sin\!\alpha  = U^{-1}.
\eqno (7)$$

By generating perpendicular electric and  magnetic  fields  we
determine empirically the directions in space. If  these  fields
satisfy the condition (6), then, according  to  the  hypothesis,
for each type of  tachyons  under consideration Eqs.~(5)
and  (7)  determine  four
variants of the complete kinematical conditions for the produced
tachyon.  The existence of four variants results  from  relations
(1)--(3) and~(7). Namely, there are double signs of  the  nonzero
components $E_{i}$ and~$H_{i}$, a double sign of~$ j$
(i.e. a double sign of
$ w$ since $jw < 0$), and $\sin\!\alpha  = \sin (\pi  - \alpha )$,
i.e. we apparently have
eight variants, but  each  one of these  three  items
depends on two others.

In order to determine the principal empiric conditions  for
the tachyon production, we should also know the quantity $ A$
and the rest
mass $M$ of the generative particle. I am able  to  estimate  only
their lower limits~[1].

In the case of $ A$, we get
$$
A \gtrsim 6.9 \times 10^{17} {\rm\ esu/cm}^{2}
{\rm\ or\ oersted.}\eqno (8)
$$

In the case of $M$, I am able to  estimate  its  lower  limit
only when $| w|  \cong  1 $ (thus for $U \cong  1$; note
that $| w|  > 1$ and $U > 1$),
i.e. when the produced tachyon is very ``slow'' in  the  proper
reference  frame  of  the  generative  particle.\footnote[2]{Such a
tachyon can, however,
be observed as considerably faster than light if the sense of its
velocity is opposite in the laboratory reference frame to the sense
of the generative particle velocity (sufficiently high but subluminal
of course); cf. remarks on the backward tachyons in Section~3.}
Laborious calculation~[1] gives
$$
M \gtrsim 75\ {\rm GeV/}c^{2}.\eqno (9)
$$

Our hypothesis concerns the production of the tachyons generating
convex spacetimes; and such tachyons can
exist autonomously. Let us call them {\it principal tachyons}. Each
principal tachyon may be accompanied with an arbitrary (formally)
number of tachyons generating
concave spacetimes. The latter
tachyons cannot exist autonomously but they can exist if they form
a ``star of tachyons'' together with a principal
tachyon. Let us call them {\it accompanying tachyons}.
All the tachyons
forming their ``star'' are born at one event (common creation point).
Details are given in Ref.~[1], and briefly in Ref.~[9].


\centerline{{{\bf\large 3. Comments on the empiric possibilities}}}


The production conditions determined by our hypothesis can occur in
high energy collisions with atomic nuclei other than the
proton. In such collisions we can locally obtain the conditions~(4)
(for details see Ref.~[1]) and the relativistic
intensification of the electromagnetic fields of nuclei necessary
to satisfy the condition~(8). It is easy to calculate that this
intensification gives $ U \cong 1$, i.e. the condition (9) holds.
Thus the gauge boson Z$^{0} $ is the lightest known candidate for
the generative particle. Though the mean life of this boson is very
short, the production conditions can be satisfied. In fact, if a
subatomic particle of sufficiently high energy strikes a nucleon
included in an atomic nucleus and produces the boson~Z$^{0} $, then
{\it in statu nascendi\/} this boson moves with respect to the
nucleus (its remainder) with a velocity that sufficiently intensifies
the electromagnetic field. In particular, neutrons present in nuclei
should be struck by neutral particles, while protons by
negatively charged ones. In the case of nuclei so large that we may
speak of peripheral nucleons, the collision with such a nucleon
(``tangent'' collision) is the most effective. Note that the
principal m-tachyon is produced {\it only\/} when the
proton in the~$ ^{2}$H,
and perhaps~$ ^{3}$H, nucleus is appropriately struck. When
designing controlled collisions, we can practically use only
electrons or antiprotons as the striking particles. In all the
mentioned collisions we have $ U \cong 1$ and therefore, by Eq.~(7),
the striking particle and the produced principal tachyon have
practically the same direction of motion, but according to our
theory they may have different senses~[1]. In the case of opposite
senses for brevity we shall be speaking about {\it backward
tachyons}, and in the case of the same senses about {\it forward
tachyons}. This nomenclature relates to the principal tachyons
only.

The collisions described above should occur in air showers and can
be realized in or at some high-energy colliders. Let us discuss
these two cases in terms of the {\it laboratory\/} (and thus the
{\it earth\/}) reference frame.

The collisions producing tachyons should occur
in the air showers
initiated by cosmic (primary) particles
of energy of $ \sim\!\!10^{13}$~eV
and greater (events above $ 10^{20}$~eV have been reported~[10]).
Thus our hypothesis justifies air shower experiments
designed to detect tachyons. The time-of-flight measurement
experiments (e.g. described in Refs. [5,11,12]) are obviously
more credible than the experiments described and/or cited in
Refs. [2--4,7,8] and designed only to detect charged particles
preceding the relativistic fronts of air showers, though a
massive-measurement experiment of the latter type performed by
Smith and Standil with the use of detector telescopes [13]
has had great
weight. Tachyon candidates were observed in the time-of-flight
experiments [5,11,12] and in many ``preceded front'' ones
including that described in Ref.~[13], but these unlucky candidates
were sunk in backgrounds and/or statistics. Thus, formally, we have
to consider the results as negative. In the light of our hypothesis,
however,
properly designed experiments with air showers (``poor man's
accelerator''~[12]) are worth repeating, the more so as they are
relatively inexpensive.

Let us note that no forward tachyons can be observed in any air
shower experiment performed in the terrestrial reference frame,
since these tachyons cannot practically precede the shower fronts.
In fact, it is easy to calculate from relations~(5), (8), and from
the relativistic law of
addition of velocities that the forward e-tachyons produced in
collisions with nuclei $ ^{40}$Ar can move in this
reference frame with speeds not greater than
$ \sim\!\!1.0000008c $. In the case of nuclei $ ^{16}$O or $ ^{14}$N, 
or $ ^{2}$H in the case of production of the forward m-tachyons,
the upper speed limit is still lower. On the other hand, some
tachyons accompanying those ``slow'' forward tachyons may travel
considerably faster than light towards the ground.
This is possible provided
that the angle, denoted by $ \psi $ for short, between the motion
directions of such a forward tachyon and of its accompanying tachyon
is sufficiently large.\footnote[3]{In every given reference
frame, if a principal tachyon moves with a speed $ |W| < \infty $
and if the
angle $ \psi $ between the velocity $ W $ and velocity $ V $ of a 
tachyon accompanying this principal one is, for simplicity, smaller
than $ \pi/2 $,
then $ |V| \leq c|W|/[c\cos\!\psi + (W^{2} - c^{2})^{1/2}\sin\!\psi] $
and there is a lower limit for
$ \psi $, namely
$ \arccos (c/|W|) < \psi < \pi/2 $ in the case under
consideration. Of course $ |V| > c $ and $ |W| > c$.} Unfortunately,
these fast accompanying tachyons cannot be observed in typical
``preceded front'' experiments since they escape from the showers
sidewise. They could be observed in the previous
time-of-flight experiments in the cases when the shower axis was
largely inclined with respect to the flight corridor of the
detector (large~$ \psi $).

The described situation seems to explain the poor statistics
obtained from the previous experiments, and suggests how to design
new air shower experiments to search for tachyons. It seems that
the best solution would be an {\it apparatus with many time-of-flight
corridors of various directions}. In order to increase efficiency,
such an apparatus should be possibly close to the region of tachyon
production (mountains? balloons?). To increase credibility, simple
air shower detectors (placed on the ground for convenience) can
additionally be used. They should be far from the main apparatus
(its projection on the ground) to act when $ \psi $ is large, i.e.,
when the registered showers are remote or largely inclined.
If some tachyon flights through the main apparatus coincide
with the signals from some of the additional detectors,
then we get stronger evidence that tachyons are produced in
air showers. The use of the main apparatus alone should also give
us valuable results without detecting any showers.

The appearance of tachyon candidates in some
previous ``preceded front''
experiments can be explained as the arrival of tachyons accompanying
the backward tachyons. The backward tachyons produced in air showers
are slightly faster than $ 5c/3$ in the terrestrial reference frame.
Thus, at sufficiently high altitudes
\underline{(balloons? satellites?)},
they should be easily identified as tachyons
\underbar{and} \underbar{the} \underbar{detecting} \underbar{system}
\underbar{can be} \underbar{\it very simple}.

Failure of the previous air shower experiments may also be explained
by the very low deuterium content (cf. the beginning of this section)
in the earth's atmosphere. Indeed, if the principal e-tachyons do
not exist in nature but the principal m-tachyons do
(cf. Footnote~1), then the probability of
production of principal tachyons is very low. Then, however, this
probability strongly depends on weather. Roughly speaking, the
\underline{cloudier the skies} the higher
the probability. \underbar{(This also concerns}
\underbar{the case of the $ 5c/3$} \underbar{backward}
\underbar{tachyons} \underbar{when} \underbar{the detecting}
\underbar{system} \underbar{must be} \underbar{of course}
\underbar{above} \underbar{the clouds.)}
It seems that \underline{the effect of cloudiness}
has not been taken into account in the
experiments performed hitherto. If the principal tachyons are
only the m-tachyons,
then the efficiency of air shower experiments may be
increased by introducing extra deuterium. For instance,
we can place the above mentioned
apparatus (i.e. that with many time-of-flight corridors)
{\it inside\/} a large balloon filled
with hydrogen and next dispatch the balloon to the region of
tachyon production.

In the case of performing tachyon search experiments with the use of
accelerators we can choose the striking particles (practically
either electrons or antiprotons), the nuclei to be struck,
and the energy of collisions. Relations (8) and (9) mean that
the strongest colliders should be employed. At present,
however, we can
only direct a beam of electrons or antiprotons onto a stationary
target. This would give us principal tachyons such as in the case of
air showers, i.e. forward tachyons so ``slow'' that
indistinguishable as tachyons and backward tachyons
slightly faster than $ 5c/3$. As regards accompanying tachyons,
we would have a much better situation since the target can be
surrounded with tachyon detectors, e.g. with time-of-flight ones.
The fact that tachyon candidates were observed in air shower
experiments indicates that there should be no problems with the
range of tachyons in the collider experiments. A collider with
a high energy beam of atomic nuclei would extend our empiric
possibilities. We could then control the observed speeds of
backward and forward tachyons and, in consequence, change the observed
velocities of the accompanying tachyons. Besides, we could then
produce principal m-tachyons (cf. the preceding paragraph), which
is impossible in the near future when a stationary target is used.
For instance, a beam of electrons of energy of $ \sim\!25$~GeV
or a beam of antiprotons of energy of $ \sim\!0.1$~TeV when
colliding with a beam of deuterons of energy of $ \sim\!\!1$~TeV 
($ \sim\!0.5$~TeV/u) or of $ \sim\!0.24$~TeV
($ \sim\!0.12$~TeV/u), respectively,
would already realize the production conditions, whereas in the case
of the deuterium target the energy of the striking negative
particles must be $ \sim\!26$~TeV. When using stationary targets
to produce principal e-tachyons, we need the striking negative
particles of energy of $ \sim\!0.8$~TeV for the targets made of
heavy nuclei, and of $ \sim\!2$~TeV for the targets made of light
nuclei.

Let us note that in the experiments designed to detect tachyons the
existence of a reference frame preferred for the tachyons should be
taken into account. In terrestrial experiments we should
therefore analyze the measurements in correlation with the time of
the day, and additionally, in long-lasting experiments, with the 
season of the year. It seems obvious that from this point of
view the experiments with the use of colliders are more suitable
than those with air showers.

The existence of the reference frame preferred for the tachyons
has been considered or postulated
by many authors. Most of the relevant literature is cited in Refs.
[14--16]. Some ideas are, however, in conflict with empiric data,
some others can only be verified by means of tachyons.
According to the latter ideas such a frame is imperceptible for
bradyons and luxons, which means that this frame is a usual
non-preferred inertial reference frame for all the tachyonless
phenomena. This is not contradictory to relativity (which
has been verified only in the bradyonic and luxonic domains) and is
not empirically ruled out since tachyons have not yet been employed.
The most natural idea (i.e. when the (local) Minkowski's spacetime
is assumed to be spatially isotropic also for tachyons) has
thoroughly been analyzed in Section~3 of Ref.~[15]. Following this
idea, many authors suggest that the frame in question is
that in which the cosmic microwave background radiation is
isotropic. If their intuition is correct, then in terrestrial
experiments this frame can be revealed only by means of tachyons
which are very fast (over $ \sim\!800c$) in the laboratory reference
frame. If, however, the ``tachyon corridor'' described by Antippa
and Everett [17,18] did exist, then ``slow'' tachyons would be 
sufficient to reveal it.

\newpage

\centerline{{{REFERENCES}}}

\vglue12pt

\noindent
[1] J.K. Kowalczy\'nski, {\it The Tachyon and its Fields}, Polish
Academy of Sciences, Warsaw 1996.

\noindent
[2] J.R. Prescott, {\it J. Phys.} {\bf G2}, 261 (1976).

\noindent
[3] L.W. Jones, {\it Rev. Mod. Phys.} {\bf 49}, 717 (1977).

\noindent
[4] P.N. Bhat, N.V. Gopalakrishnan, S.K. Gupta, S.C. Tonwar,
{\it J. Phys.} {\bf G5}, L13 (1979).

\noindent
[5] A. Marini, I. Peruzzi, M. Piccolo, F. Ronga, D.M. Chew, R.P. Ely,
T.P. Pun, V. Vuillemin, R. Fries, B. Gobbi, W. Guryn, D.H. Miller,
M.C. Ross, D. Besset, S.J. Freedman, A.M. Litke, J. Napolitano,
T.C. Wang, F.A. Harris, I. Karliner, Sh. Parker, D.E. Yount,
{\it Phys. Rev.} {\bf D26}, 1777 (1982).

\noindent
[6] Particle Data Group, Review of Particle Properties,
{\it Phys. Rev.} {\bf D50} (1994) No. 3--I, p. 1811.

\noindent
[7] R.W. Clay, {\it Aust. J. Phys.} {\bf 41}, 93 (1988).

\noindent
[8] R.W. Clay, P.C. Crouch, {\it Nature} {\bf 248}, 28 (1974).

\noindent
[9] J.K. Kowalczy\'nski, {\it Phys. Lett.} {\bf 74A}, 157 (1979).

\noindent
[10] T.K. Gaisser, T. Stanev, {\it Phys. Rev.} {\bf D54}, 122 (1996).

\noindent
[11] F. Ashton, H.J. Edwards, G.N. Kelly, {\it Nucl. Instrum.
Methods} {\bf 93}, 349 (1971).

\noindent
[12] H. H\"anni, E. Hugentobler, in {\it Tachyons, Monopoles, and
Related Topics}, ed. E. Recami, North-Holland, Amsterdam
1978, p. 61.

\noindent
[13] G.R. Smith, S. Standil, {\it Can. J. Phys.} {\bf 55},
1280 (1977).

\noindent
[14] R. Girard, L. Marchildon, {\it Found. Phys.} {\bf 14}, 535 (1984).

\noindent
[15] J.K. Kowalczy\'nski, {\it Int. J. Theor. Phys.} {\bf 23},
27 (1984).

\noindent
[16] J. Rembieli\'nski, {\it Int. J. Mod. Phys.} {\bf A12},
1677 (1997).

\noindent
[17] A.F. Antippa, A.E. Everett, {\it Phys. Rev.} {\bf D8},
2352 (1973).

\noindent
[18] A.F. Antippa, {\it Phys. Rev.} {\bf D11}, 724 (1975).

\end{document}